\documentclass[preprint,12pt]{elsarticle}%
\usepackage{amssymb}
\usepackage{amsfonts}
\usepackage{amsmath}
\usepackage{graphicx}%
\providecommand{\U}[1]{\protect\rule{.1in}{.1in}}
\graphicspath{{E:/dottorato/Coulomb_damper/Testo/Paper/Immagini/}}
\journal{journal}

\begin{document}
%
%TCIMACRO{\TeXButton{Begin frontmatter}{\begin{frontmatter}}}%
%BeginExpansion
\begin{frontmatter}%
%EndExpansion



%TCIMACRO{\TeXButton{Title}{\title
%{Improved Muller approximate solution of the pull-off of a sphere from a viscoelastic substrate}%
%}}%
%BeginExpansion
\title
{Improved Muller approximate solution of the pull-off of a sphere from a viscoelastic substrate}%
%EndExpansion


%TCIMACRO{\TeXButton{Author}{\author{M. Ciavarella}}}%
%BeginExpansion
\author{M. Ciavarella}%
%EndExpansion
%

%TCIMACRO{\TeXButton{Address}{\address
%{Politecnico di BARI. DMMM department. Viale Gentile 182, 70126 Bari. Mciava@poliba.it}%
%}}%
%BeginExpansion
\address
{Politecnico di BARI. DMMM department. Viale Gentile 182, 70126 Bari. Mciava@poliba.it}%
%EndExpansion
%

%TCIMACRO{\TeXButton{Begin abstract}{\begin{abstract}}}%
%BeginExpansion
\begin{abstract}%
%EndExpansion

The detachment of a sphere from a viscoelastic substrate is clearly a
fundamental problem. In the case viscoelastic dissipation is concentrated at
the contact edge, and the work of adhesion follows a quite popular simplified
model, Muller has suggested an approximate solution, which however is based on
an empirical observation. We revisit Muller's solution and show it leads to
very poor fitting of the actual full numerical results, particularly for the
radius of contact at pull-off, and we suggest an improved fitting of the
pull-off which works extremely well over a very wide range of withdrawing
speeds, and correctly converges to the JKR value at very low speeds.%

%TCIMACRO{\TeXButton{End abstract}{\end{abstract}}}%
%BeginExpansion
\end{abstract}%
%EndExpansion
%

%TCIMACRO{\TeXButton{Begin keyword(s)}{\begin{keyword}}}%
%BeginExpansion
\begin{keyword}%
%EndExpansion

Viscoelasticity, Adhesion, JKR theory, soft matter%

%TCIMACRO{\TeXButton{End keyword(s)}{\end{keyword}}}%
%BeginExpansion
\end{keyword}%
%EndExpansion
%

%TCIMACRO{\TeXButton{End frontmatter}{\end{frontmatter}}}%
%BeginExpansion
\end{frontmatter}%
%EndExpansion



\section{\bigskip Introduction}

The problem of viscoelastic dissipation during crack growth or contact peeling
has attracted much interest due to its fundamental importance in many areas of
science and technology. Many authors have applied fracture mechanics concepts
and made extensive measurements (Gent and Schultz, 1972, Barquins and Maugis
1981, Gent, 1996, Gent \& Petrich 1969, Andrews \& Kinloch, 1974, Barber
\textit{et al}, 1989, Greenwood \& Johnson, 1981, Maugis \& Barquins, 1980,
Persson \& Brener, 2005) postulating peeling involves an effective work of
adhesion $w$ as the product of adiabatic value $w_{0}$ and a function of
velocity of peeling of the contact/crack line and temperature, namely%
\begin{equation}
w=w_{0}\left[  1+k\left(  a_{T}v_{p}\right)  ^{n}\right]  \label{wvisco}%
\end{equation}
where $k,n$ are constants of the material, with $n$ in the range $0.1-0.8$ and
$a_{T}$ is the WLF factor (Williams, Landel \& Ferry, 1955) which permits to
translate results at various temperatures $T$ from measurement at a certain
standard temperature. The details of the derivation from crack models
involving cohesive Barenblatt zones or models "truncating" or "blunting" crack
tip dissipation (Barber Donley and Langer 1989, Greenwood and Johnson, 1981,
Persson \&\ Brener, 2005) vary, but the form (\ref{wvisco}) remains the most
popular simple choice, and therefore a baseline for comprehension of possible
mechanics of contact and crack problems.

In the case of adhesive contact of the fundamental spherical geometry, various
authors (Barquins \& Maugis, 1981, Greenwood \&\ Johnson, 1981, Muller, 1999)
have attempted to apply the fracture mechanics formulation with the model
(\ref{wvisco}), and some approximate results have been given in terms of
explicit dependences of the pull off force or work, contact radius and
approach at pull-off see ref. (Muller, 1999), which we shall revisit here in
comparison with full numerical simulation, finding very significant
discrepancies, and suggesting some improved fitting of the numerical results,
at least for the pull-off force which is the quantity of greater interest.

\section{Spherical contact mechanics theory}

The fracture mechanics formulation for the adhesive contact problem for a
sphere is classic, and we shall revisit here only the essentials. \bigskip We
consider the stress intensity factor at the contact edge is due to the
difference between $P_{1}$, the load required to maintain a contact radius $a$
in the absence of adhesion%
\begin{equation}
P_{1}\left(  a\right)  =\frac{4}{3}\frac{E^{\ast}}{R}a^{3} \label{P1}%
\end{equation}
where $E^{\ast}=E/\left(  1-\nu^{2}\right)  $ is the plane strain elastic
modulus ($E$ being Young's modulus and $\nu$ Poisson's ratio) and $P$ , the
smaller load to maintain the same contact radius in the presence of adhesion.
So we find the strain energy release rate as\footnote{The factor 2 which is
missing in Muller (1999) comes from the fact that strain energy exists only in
one material, assuming the other is rigid. For two identical materials,
$\frac{1}{E^{\ast}}=\frac{2}{E_{1}^{\ast}}$ and we return to the standard LEFM
case with $G\left(  a\right)  =\frac{K\left(  a\right)  ^{2}}{E_{1}^{\ast}}$.}%
\begin{equation}
G\left(  a,P\right)  =\frac{K\left(  a,P\right)  ^{2}}{2E^{\ast}}%
=\frac{\left(  P_{1}\left(  a\right)  -P\right)  ^{2}}{8\pi E^{\ast}a^{3}}
\label{G1}%
\end{equation}

In the adhesionless conditions, the remote approach is $\alpha_{1}\left(
a\right)  =\frac{a^{2}}{R}$, so in the adhesive condition we have to decrease
this by an amount given by a flat punch displacement $\Delta\alpha=\frac
{P_{1}-P}{2E^{\ast}a}$ (since in moving from the adhesionless to the adhesive
solution we keep the contact area constant) giving the general result for
approach
\begin{equation}
\alpha\left(  a,P\right)  =\frac{a^{2}}{R}-\frac{P_{1}\left(  a\right)
-P}{2E^{\ast}a} \label{alfa}%
\end{equation}
from which we can obtain $P\left(  a,\alpha\right)  $ using (\ref{P1})
\begin{align}
P\left(  a,\alpha\right)   &  =P_{1}\left(  a\right)  +2E^{\ast}a\alpha\left(
a,P\right)  -2E^{\ast}\frac{a^{3}}{R}\nonumber\\
&  =\frac{2E^{\ast}a}{R}\left(  R\alpha\left(  a,P\right)  -\frac{a^{2}}%
{3}\right)  \label{P}%
\end{align}
which corresponds to Muller (1999) equation 10, whereas using (\ref{G1})%
\begin{equation}
G\left(  a\right)  =\frac{\left(  P_{1}\left(  a\right)  -P\right)  ^{2}}{8\pi
E^{\ast}a^{3}}=\frac{E^{\ast}}{2\pi aR^{2}}\left(  R\alpha\left(  a\right)
-a^{2}\right)  ^{2} \label{G2}%
\end{equation}
which corresponds to Muller (1999) equation 15 except for a factor 2 misprint.
For the elastic case, JKR (Johnson, Kendall \&\ Roberts, 1971) theory is
obtained by using (\ref{G2}) and (\ref{alfa})%
\begin{equation}
P=\frac{4}{3}\frac{E^{\ast}}{R}a^{3}-\sqrt{8\pi w_{0}E^{\ast}a^{3}}
\label{JKR}%
\end{equation}
Putting
\begin{equation}
\zeta=\left(  \frac{\pi w_{0}}{6RE^{\ast}}\right)  ^{1/3} \label{zita}%
\end{equation}
we have at $P=0$ from (\ref{JKR}) and (\ref{P})%
\begin{align}
a_{0}  &  =\left(  \frac{9}{2}\pi R^{2}\frac{w_{0}}{E^{\ast}}\right)
^{1/3}=3R\zeta\label{a0}\\
\alpha_{0}  &  =\frac{a_{0}^{2}}{3R}=3R\zeta^{2} \label{alfa0}%
\end{align}
where there is a factor 3 misprint in Muller (1999) equation 19.

\section{Viscoelasticity}

Now, for a viscoelastic material, the material dissipation at the crack
tip/contact edge requires that energy balance imposes the velocity of crack
according to (\ref{wvisco}). Further, we can write the velocity of the contact
edge as
\begin{equation}
v_{p}=-\frac{da}{dt}=v\frac{da}{d\alpha} \label{vp}%
\end{equation}
where $v$ is the remote pull-off rate imposed by the loading equipment. The
condition $G\left(  a\right)  =w$ therefore defines a differential equation
for $a=a\left(  \alpha\right)  $ obtained using (\ref{G2}, \ref{vp})
\begin{equation}
\frac{1}{k^{1/n}a_{T}v}\left[  \frac{E^{\ast}}{2\pi aR^{2}w_{0}}\left(
R\alpha\left(  a\right)  -a^{2}\right)  ^{2}-1\right]  ^{1/n}=\frac
{da}{d\alpha} \label{diff1}%
\end{equation}

By using we the JKR\ values at zero load (\ref{a0},\ref{alfa0}) and the JKR
values for pull-off for $P_{0}=\frac{3}{2}\pi Rw_{0}$, and finally the
adiabatic work of adhesion for $G$, we obtain the dimensionless variables
\begin{equation}
G^{\prime}=\frac{G}{w_{0}};\qquad P^{\prime}=\frac{P}{P_{0}};\qquad a^{\prime
}=\frac{a}{a_{0}}\text{;}\qquad\alpha^{\prime}=\frac{\alpha}{\alpha_{0}}%
\end{equation}

If we now remove the (') for simplicity in the following equation s, we
rewrite (\ref{diff1}) as
\begin{equation}
\frac{da}{d\alpha}=\beta^{-1}\left[  a^{3}\left(  \frac{\alpha}{3a^{2}%
}-1\right)  ^{2}-\frac{4}{9}\right]  ^{1/n} \label{diff2}%
\end{equation}
where we have introduced the only dimensionless factor in the problem, apart
from $n$, namely%
\begin{equation}
\beta=\left(  \frac{6RE^{\ast}}{\pi w_{0}}\right)  ^{1/3}\left(  \frac{4k}%
{9}\right)  ^{1/n}a_{T}v
\end{equation}

The latter two equation s correspond to Muller (1999) equation 24,23. The
differential equation (\ref{diff2}) can be solved for initial conditions
starting from a point on the loading curve\footnote{Strictly speaking, during
loading adhesion is reduced with respect to the adiabatic value at zero speed,
but we neglect this effect, or else we consider that loading occurs near
thermodynamic equilibrium.}, which is the JKR curve which in this
dimensionless notation and in parametric form is
\begin{equation}
P\left(  a\right)  =4\left(  a^{3}-a^{3/2}\right)
\end{equation}
and
\begin{equation}
\alpha\left(  a\right)  =3a^{2}-2a^{1/2}%
\end{equation}

After $a\left(  \alpha\right)  $ is obtained, we can compute the load which in
dimensionless form is obtained from%

\begin{equation}
P\left(  a,\alpha\right)  =2a\left(  \alpha-a^{2}\right)
\end{equation}
Notice that the strain energy release rate in dimensionless form is
\begin{equation}
G=\frac{9}{4}a^{3}\left(  \frac{\alpha}{3a^{2}}-1\right)  ^{2}%
\end{equation}

\subsection{Muller's approximate solution}

Muller (1999) in searching for the pull-off as the minimum of the $P\left(
\alpha\right)  $ curve, postulates that this is close to the minimum of
$P\left(  \alpha\right)  +G\left(  \alpha\right)  $ which is also $0$ in the
minimum. There is no fundamental reason for this mix of the dimensionless load
with the dimensionless strain energy release rate to have any special
property, and indeed we found the two minima are not necessarily very close.
Muller's postulate anyway leads to radius of contact, approach and load at
pull-off,
\begin{align}
a_{m}  &  =\kappa\beta^{q}\label{am}\\
\alpha_{m}  &  =-\kappa^{2}\beta^{2q}\label{alfam}\\
P_{m}  &  =\left\vert P_{\min}\right\vert =4\kappa^{3}\beta^{3q} \label{Pm}%
\end{align}
where $q=n/\left(  n+3\right)  $ and $\kappa=\left(  \frac{9/16}{4^{n}%
}\right)  ^{1/\left(  n+3\right)  }$ . Notice obviously that this result at
zero velocity would give \textit{incorrect results} as all values go to zero,
rather than the asymptotic values of JKR theory for thermodynamic equilibrium.

\bigskip Remark that the actual velocity of the crack line (recall $a$ and
$\alpha$ are dimensionless here, not to be confused with equation 11)
\begin{equation}
\frac{v_{p}}{v}=\frac{1}{\zeta}\left(  \frac{da}{d\alpha}\right)  _{m}%
=\frac{1}{\zeta}\frac{1}{4a_{m}}%
\end{equation}
and given $a_{m}\sim1$ while $\zeta<<1$, it is clear that $\frac{v_{p}}{v}>>1$
so that the velocity at the contact line can be much greater than the
cross-head remote velocity, which permits to make the approximation that the
bulk may be essentially in a relaxed elastic state. Notice however that, in
concentrating the effect of dissipation at the crack tip, despite the
dissipation can occur very far from it, there is another possible
approximation: indeed, the form of solution we are using is unlikely to be
reliable at extremely high speeds anyway, also for thermal effects and other
possible physical factors.

\section{Numerical results and fittings}

Here we report some results of the numerical solution of the differential
equation , comparison with Muller's approximate solution, and some improved
fitting results for the pull-off, which is (perhaps) the most important quantity.

From Fig.1 we see the withdrawing curves for an example case of low $n=0.25$,
and (b) an example showing that initial conditions seem to very weakly affect
the actual pull-off, as Muller had remarked. From Fig.2 we see that the
contact radius at pull-off is very poorly predicted by Muller's approximate
solution (\ref{am}), and it is much more weakly dependent on $\beta$. In
particular, at high $\beta$, Muller's solution predicts very large $a_{m}$
which do not make much sense. Indeed, as we have seen there is not much
dependence on the initial condition, we expect $a_{m}<1$ as when we are
unloading from equilibrium condition at zero load, and since we expect the
radius to further decrease, a fortiori we obviously end up with a smaller
radius that at zero load, which is $a_{i}=1$. An exception, where we see
$a_{m}>1$ but not by a large factor, is when there is some weak dependence on
initial conditions and we start from very high loads (see example of $P_{i}=5$
of fig.2a,c). At low $\beta$, Muller's prediction underestimates the radius at
pull-off, particularly at high $\beta$.

Also not very good predictions, but perhaps better than for contact radius,
are those for the approach at pull-off (fig.3). Here, the actual results tend
to be higher than Muller's prediction (\ref{alfam}), at all speeds, and start
off with a value near $\alpha_{m}=-0.5$ rather than from 0.

\begin{center}%
\begin{tabular}
[c]{l}%
\centering\includegraphics[height=65mm]{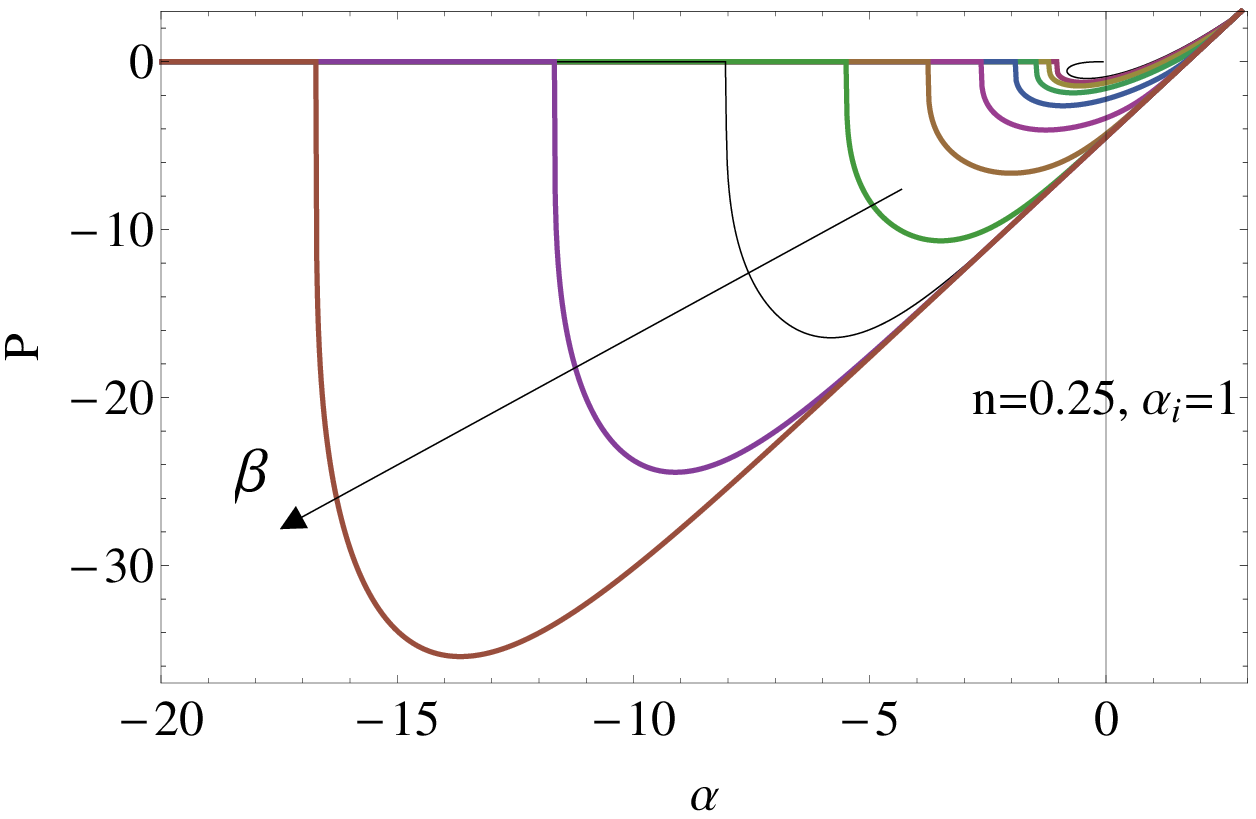}(a)
\end{tabular}

\begin{tabular}
[c]{l}%
\centering\includegraphics[height=65mm]{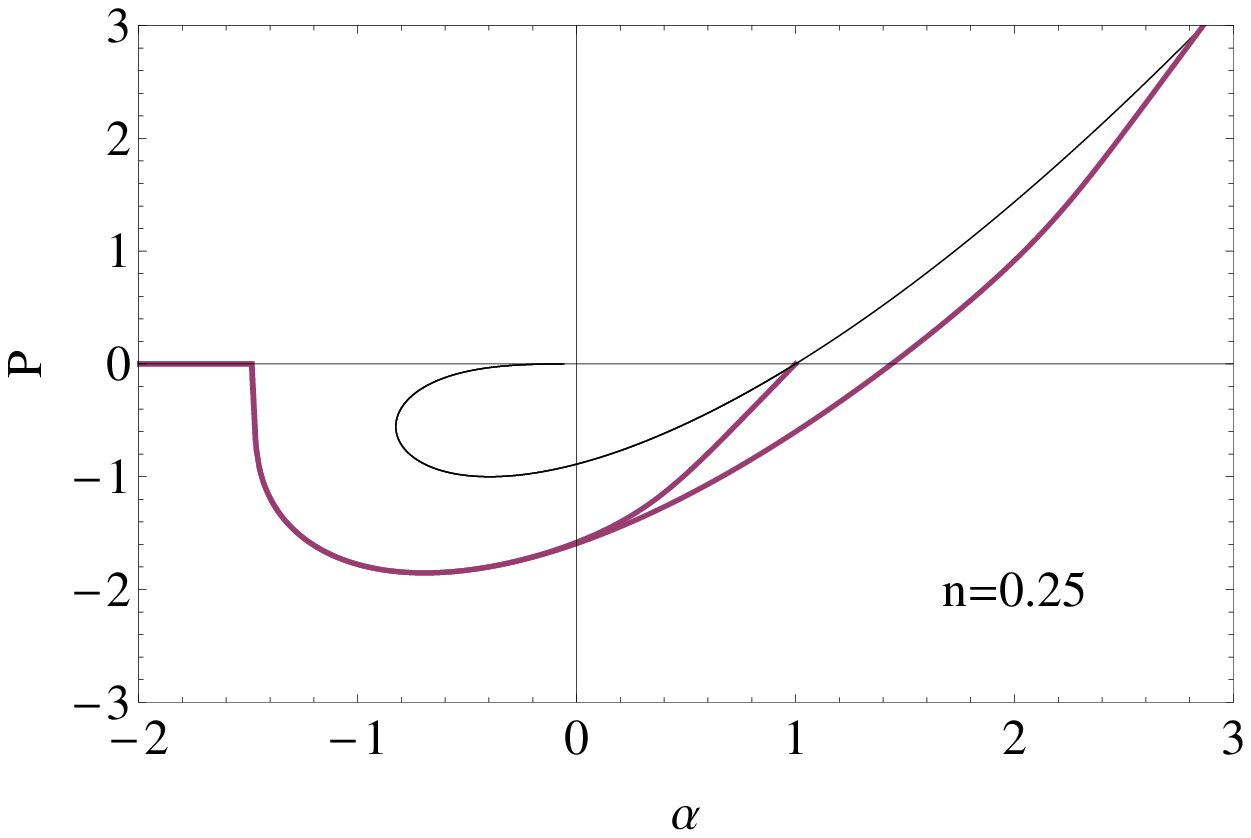}(b)
\end{tabular}

Fig.1 - Dimensionless load $P$ dimensionless approach $\alpha$ (a) for various
$\beta=2\times10^{-5}\ast15^{i}$, $\left(  i=1,10\right)  $ and for $n=0.25.$
The inner black curve is the adiabatic JKR curve. (b) very weak dependence of
pull-off on initial conditions (initial load $P=0,5$) for an example case
$\beta=0.0675$%

\begin{tabular}
[c]{ll}%
\centering\includegraphics[height=45mm]{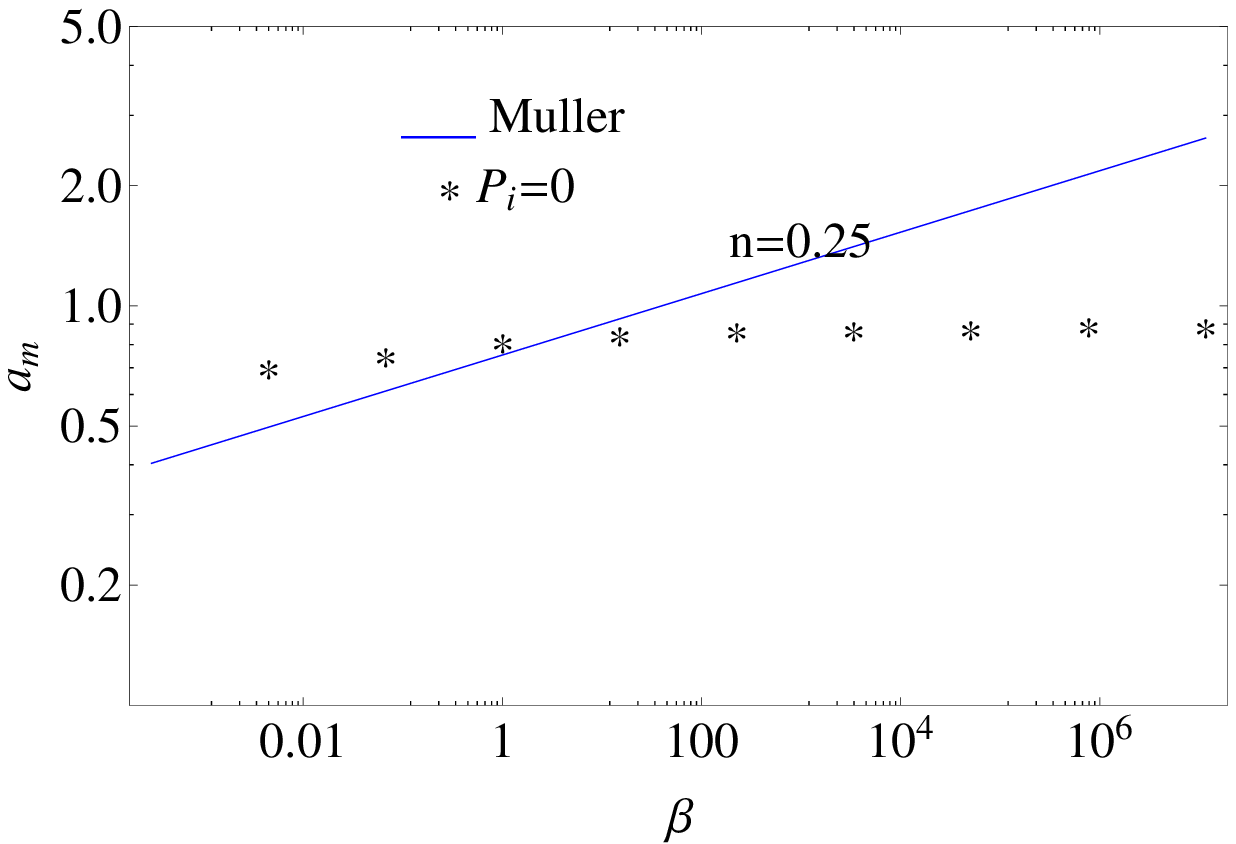}(a) &
\centering\includegraphics[height=45mm]{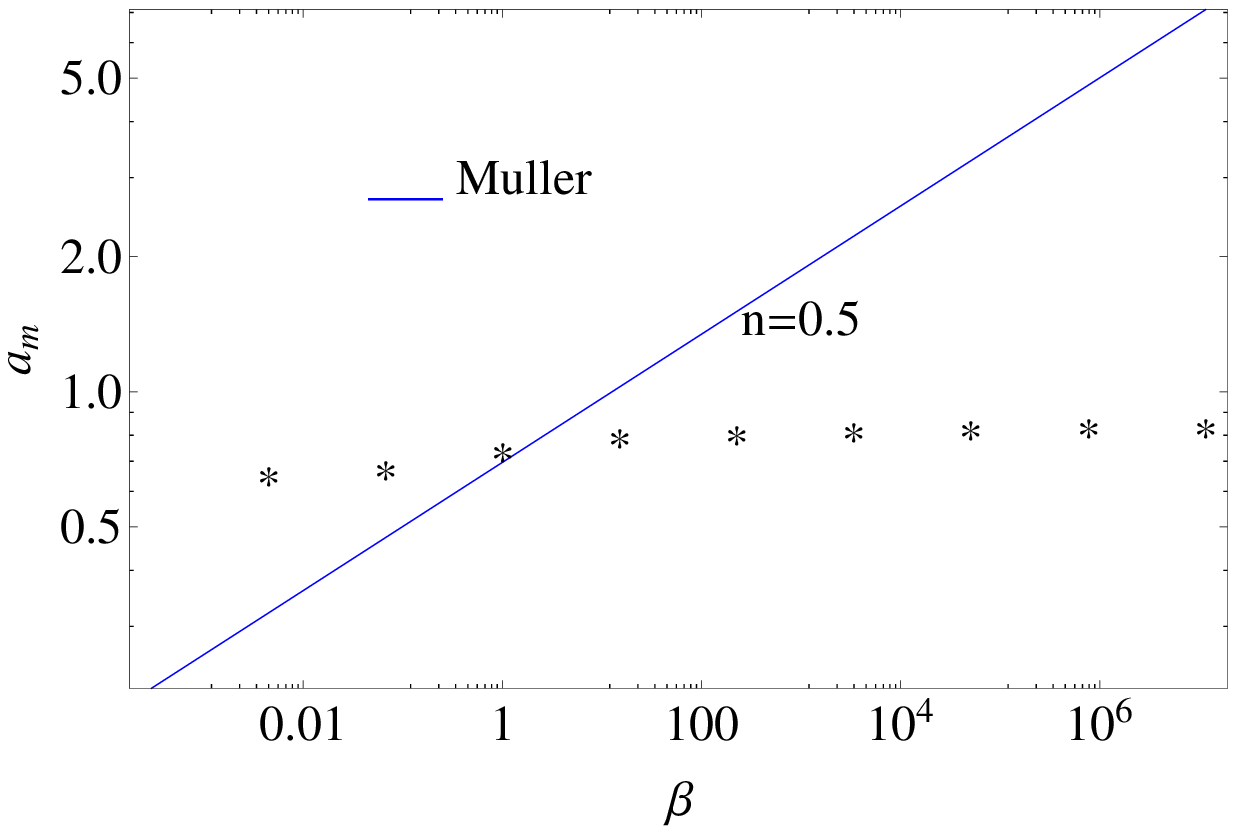}(b)
\end{tabular}

\begin{tabular}
[c]{l}%
\centering\includegraphics[height=45mm]{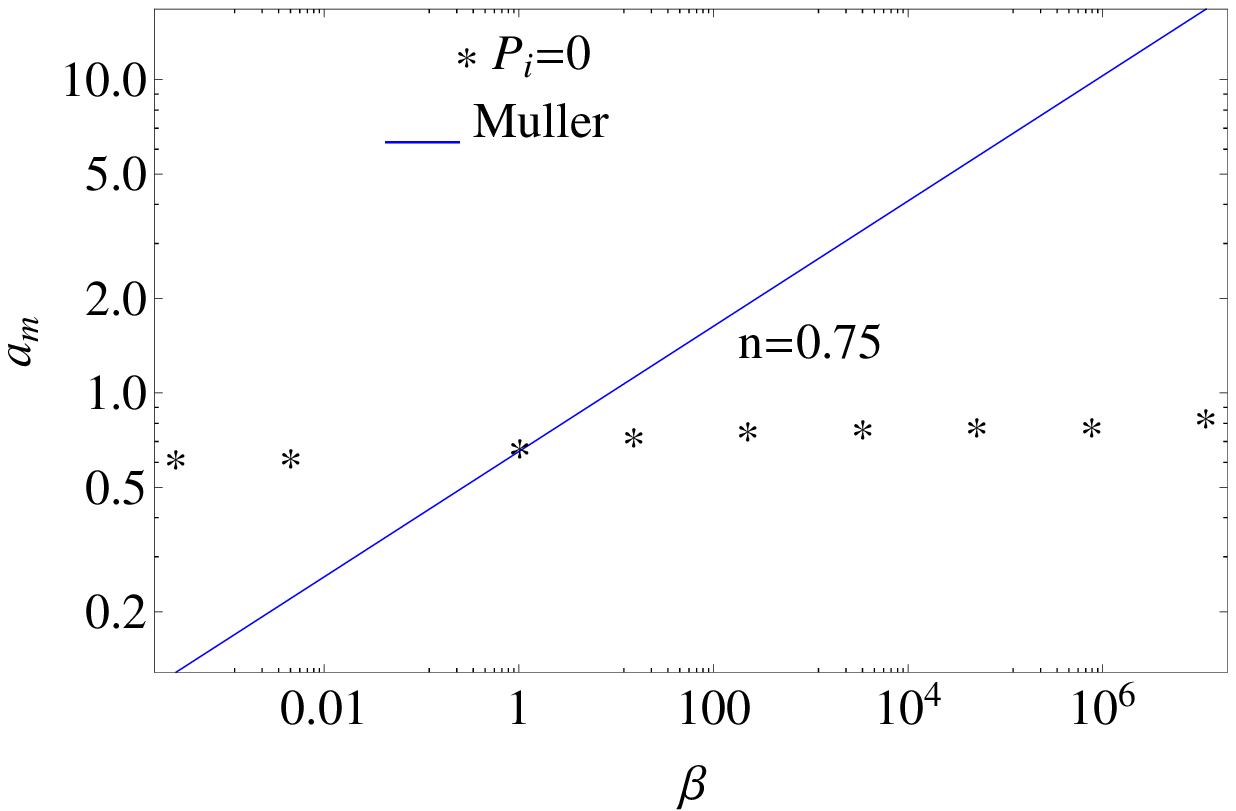}(c)
\end{tabular}

Fig.2 - Dimensionless contact radius at pull-off $a_{m}$ for $n=0.25$ (a)
$n=0.5$ (b), $n=0.75$ (c) as a function of the dimensionless speed factor
$\beta.$ (initial load in the figure $P=0$ or $5$)%

\begin{tabular}
[c]{ll}%
\centering\includegraphics[height=45mm]{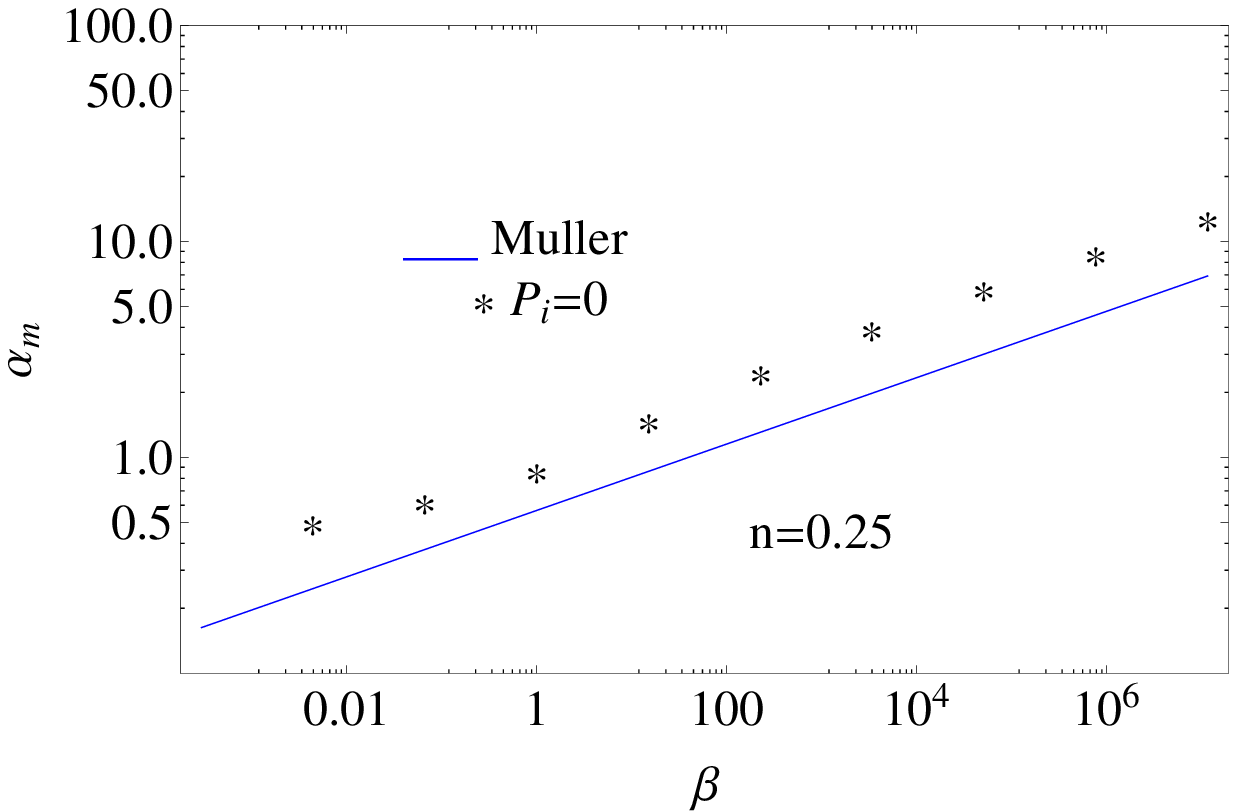}(a) &
\centering\includegraphics[height=45mm]{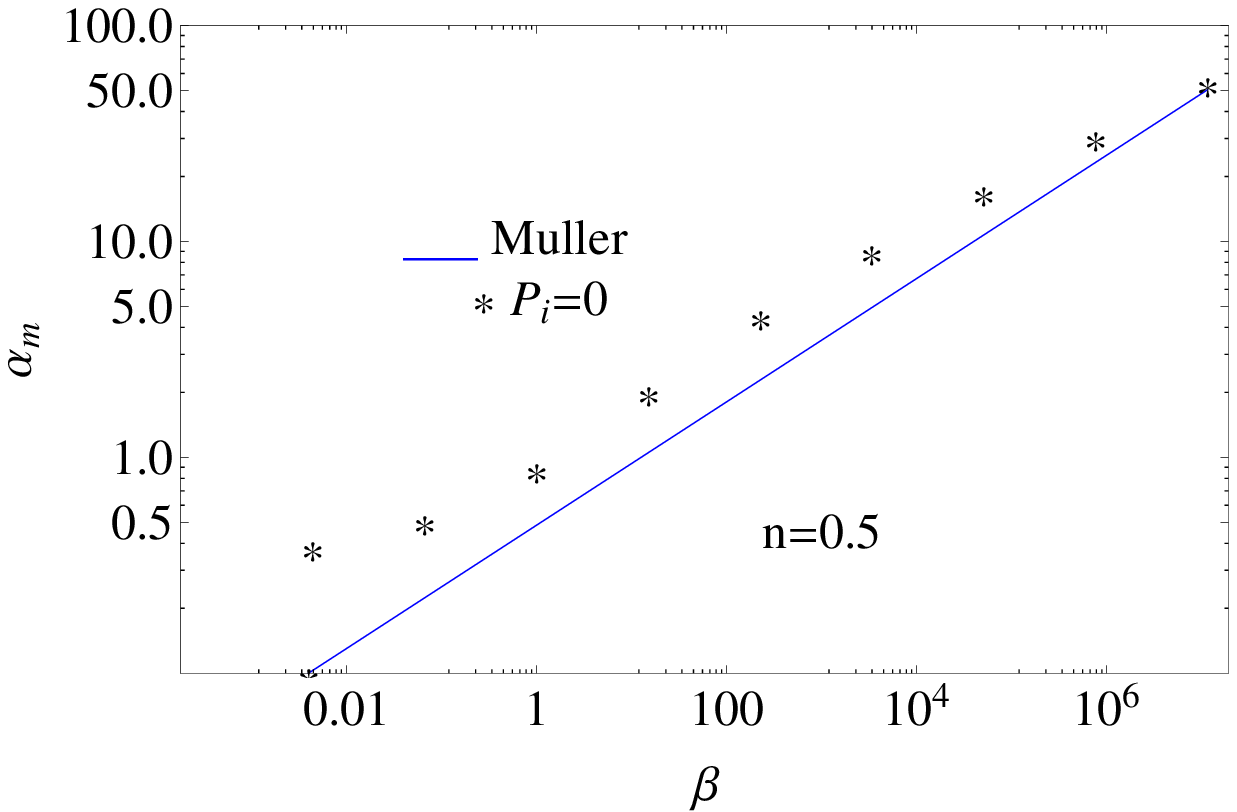}(b)
\end{tabular}

\begin{tabular}
[c]{l}%
\centering\includegraphics[height=45mm]{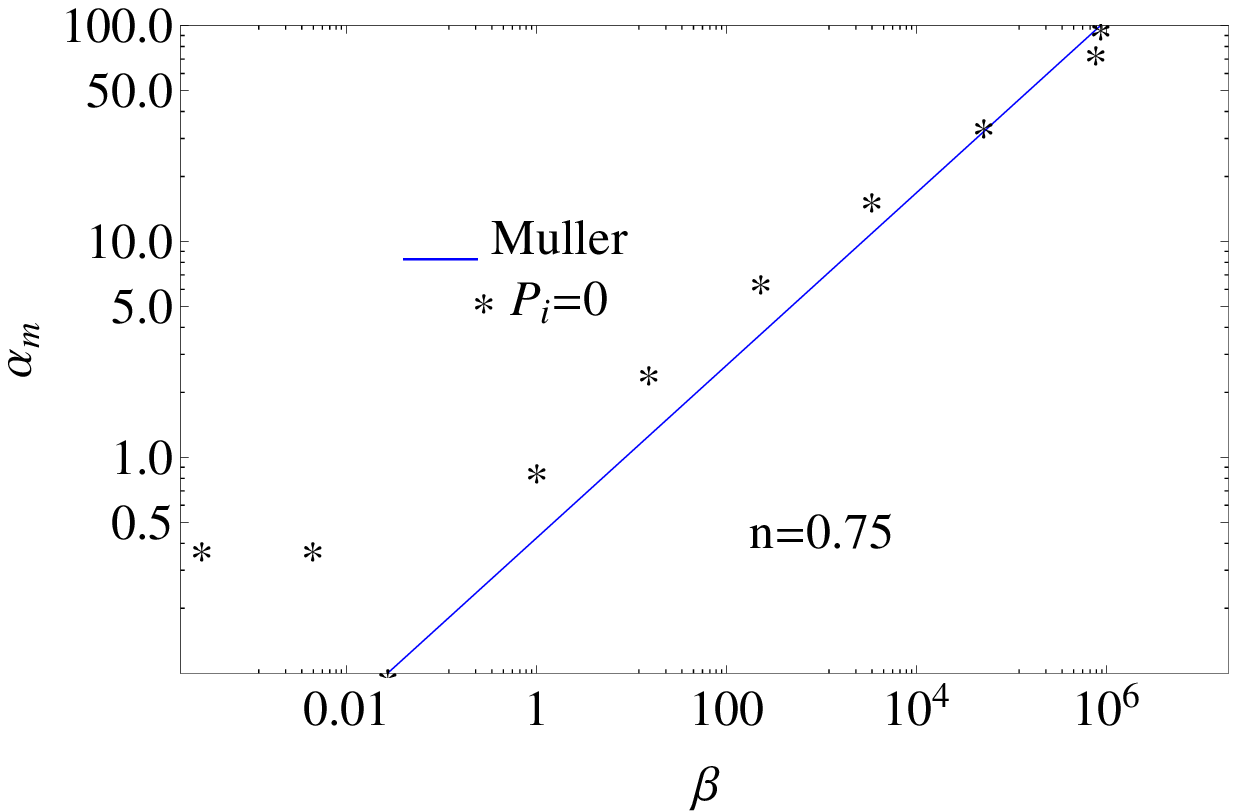}(c)
\end{tabular}

Fig.3 - Dimensionless absolute value of approach at pull-off $\left\vert
\alpha_{m}\right\vert $ for $n=0.25$ (a) $n=0.5$ (b), $n=0.75$ (c) as a
function of the dimensionless speed factor $\beta.$ (initial load in the
figure $P=0$ or $5$)
\end{center}

Considering these poor performances on $a_{m}$ and $\alpha_{m}$, the results
for the pull-off load vs Muller's prediction (see Fig.4) are relatively good
(blue line vs the markers of the numerical simulations), which is probably why
he was satisfied in his paragraph "comparison with exact calculation" where he
has only comparison with pull-off load or work for pull-off, but still we find
them only rough "estimates". It is easy to obtain much better fit of the
results, considering we have only two independent dimensionless parameters,
$n$ and $\beta$ of course, so we improve Muller's prediction in two respects:

1) we add a crossover towards the JKR value $P=1$, by adding "1" to Muller's
equation (\ref{Pm}) the JKR load;

2) we improve the power law exponent at large $\beta$ with a corrective factor
to Muller's equation (\ref{Pm}) in the form
\begin{equation}
P_{m}=\left\vert P_{\min}\right\vert =1+4\kappa^{3}\beta^{3q/c\left(
n\right)  } \label{proposal}%
\end{equation}
where
\begin{equation}
c\left(  n\right)  =1.1+n/1.65
\end{equation}
This improvement shows clearly a much better fit with respect to detailed
numerical calculations in the entire range of realistic values for $n$ and of
$\beta$ covering 10 orders of magnitude in $\beta$ which is probably more than
enough considering the other approximations made in the model, namely the form
of the work of adhesion, that there is no viscoelasticity in the bulk, no
thermal effects, and so on.

Notice that Violano and Afferrante (2019) have numerically solved the Muller
equation s, and found good correlation with experimental results. This
suggests that our solution would be very valuable for an analytical fitting of
experiments such as those of Violano \&\ Afferrante (2019).

\begin{center}%
\begin{tabular}
[c]{ll}%
\centering\includegraphics[height=45mm]{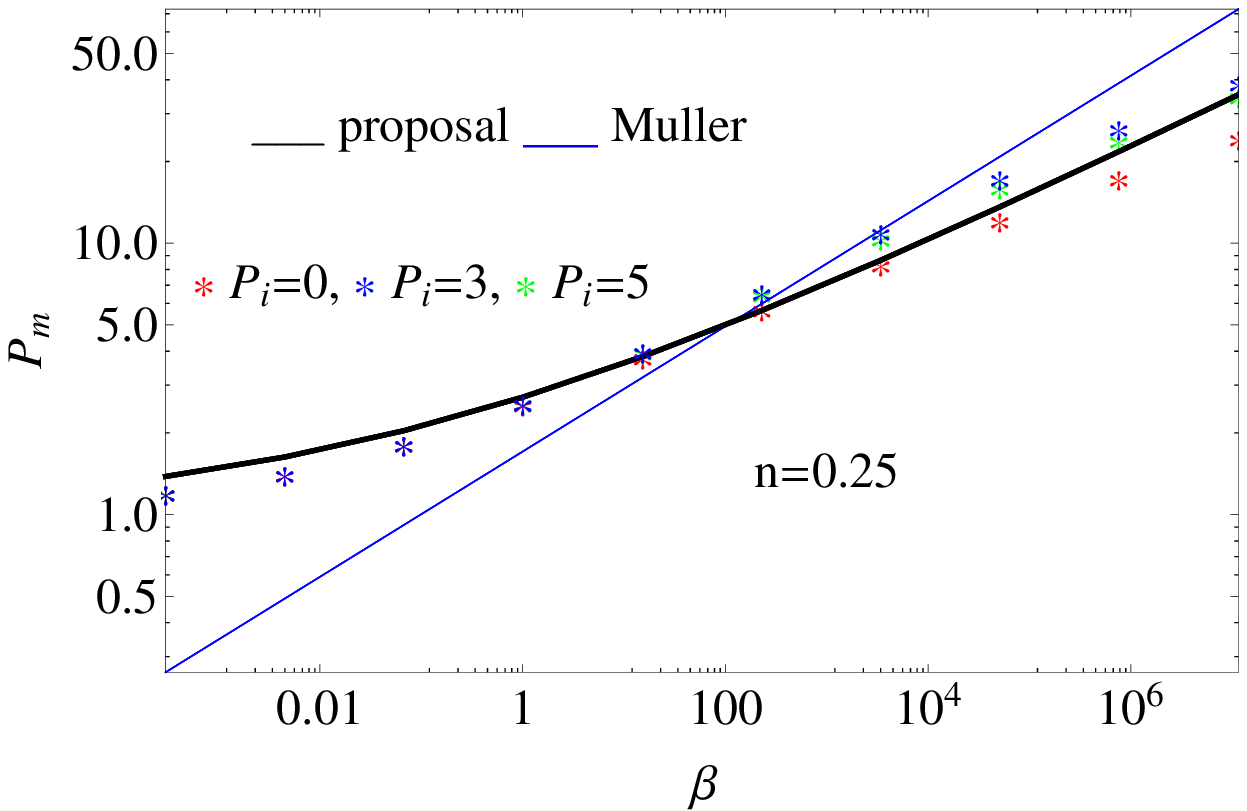}(a) &
\centering\includegraphics[height=45mm]{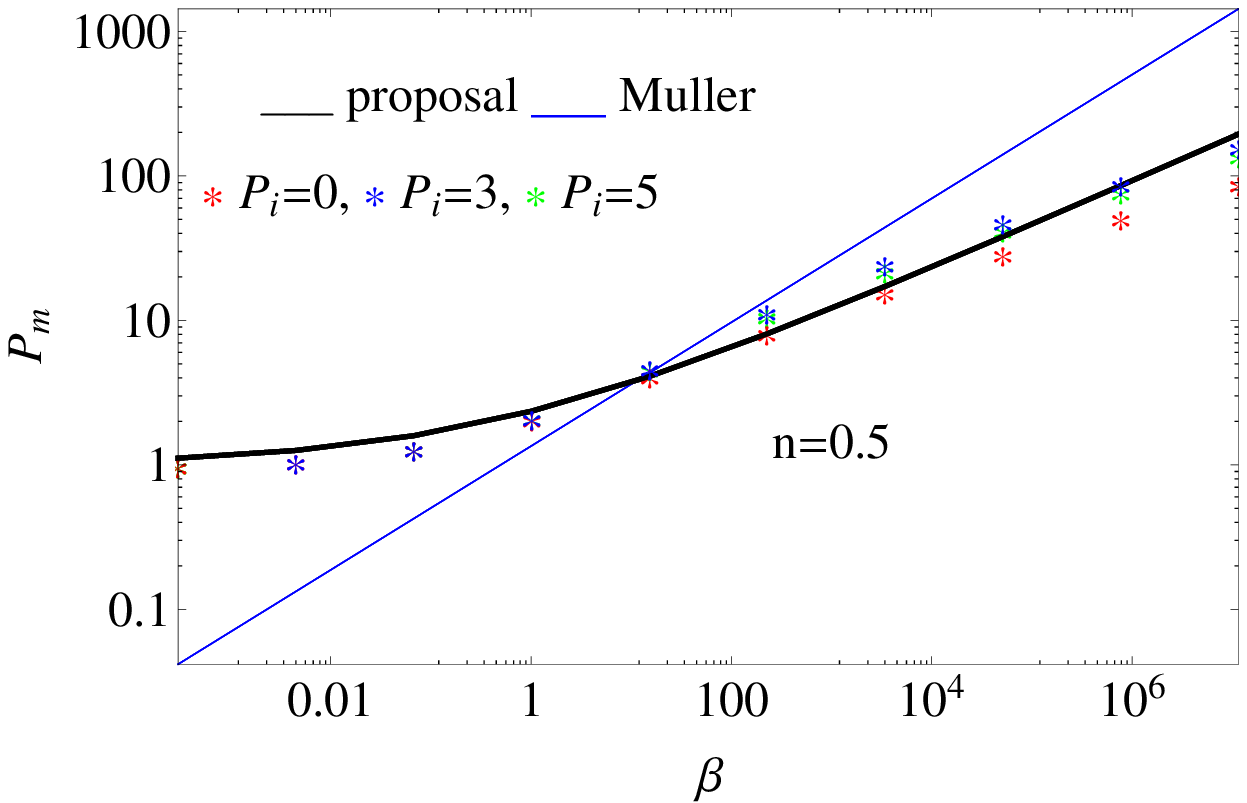}(b)
\end{tabular}

\begin{tabular}
[c]{l}%
\centering\includegraphics[height=45mm]{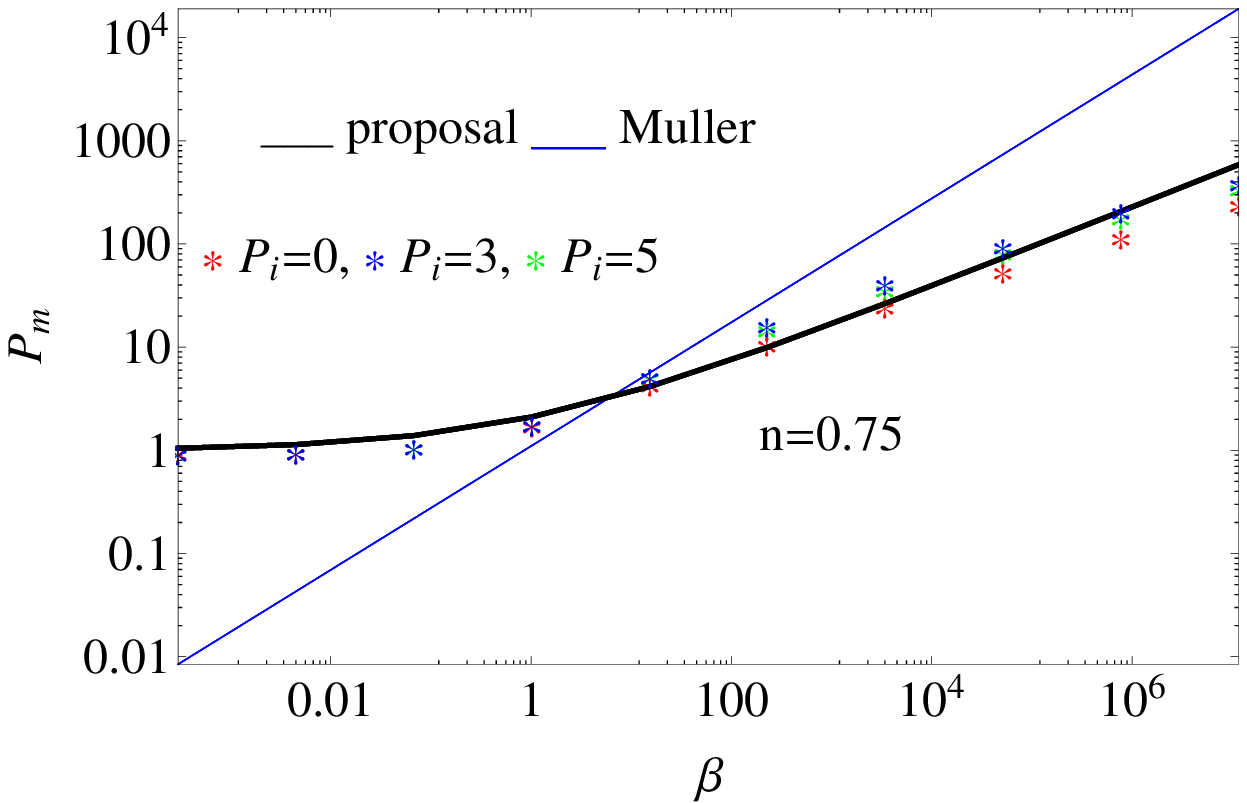}(c)
\end{tabular}

Fig.4 - Absolute value of the dimensionless load at pull off $P_{m}$ for
$n=0.25$ (a) $n=0.5$ (b), $n=0.75$ (c) as a function of the dimensionless
speed factor $\beta.$ (initial load as indicated by different colors in the
markers in the figure $P_{i}=0,3$,$5$). Blue power law curve is the Muller
(1999)\ prediction (\ref{Pm}), while the thick black solid line is our
proposal (\ref{proposal}). \bigskip
\end{center}

\section{\bigskip Conclusions}

We have revisited the Muller approximate solution for the pull-off of sphere
from a flat viscoelastic material, finding significant errors in the
approximate solution, which stem from the rather arbitrary assumption that the
pull-off condition occurs when the sum of a dimensionless load and a
dimensionless strain energy release rate has a minimum. We have added a
"cross-over" towards the JKR solution for very low velocities, and corrected
the power law enhancement of pull-off with velocity of withdrawal. The
solution can be useful for quick estimates of the effect of viscoelasticity on
the increase of adhesion in spherical geometries.

\section{Acknowledgements}

MC acknowledges support from the Italian Ministry of Education, University and
Research (MIUR) under the program "Departments of Excellence" (L.232/2016).

\section{References}

Andrews, E. H., \& Kinloch, A. J. (1974). Mechanics of elastomeric adhesion.
In Journal of Polymer Science: Polymer Symposia (Vol. 46, No. 1, pp. 1-14).
New York: Wiley Subscription Services, Inc., A Wiley Company.

Barber, M., Donley, J., \& Langer, J. S. (1989). Steady-state propagation of a
crack in a viscoelastic strip. \textit{Physical Review A}, 40(1), 366.

Barquins, M., \& Maugis, D. (1981). Tackiness of elastomers. \textit{The
Journal of Adhesion}, 13(1), 53-65.).

\bigskip Dahlquist, C. A. in Treatise on Adhesion and Adhesives, R. L. Patrick
(ed.), Dekker, New York, (1969a), 2, 219

Dahlquist, C., Tack, in Adhesion Fundamentals and Practice. (1969b), Gordon
and Breach: New York. p. 143-151.

Gent, A. N., \& Schultz, J. (1972). Effect of wetting liquids on the strength
of adhesion of viscoelastic material. \textit{The Journal of Adhesion}, 3(4), 281-294.

Gent, A. N., \& Petrich, R. P. (1969). Adhesion of viscoelastic materials to
rigid substrates. \textit{Proceedings of the Royal Society of London. A.
Mathematical and Physical Sciences}, 310(1502), 433-448.

\bigskip Greenwood, J. A., \& Johnson, K. L. (1981). The mechanics of adhesion
of viscoelastic solids. \textit{Philosophical Magazine A}, 43(3), 697-711.

Johnson, K.L. , Kendall, K. , Roberts, A.D., (1971) Surface energy and the
contact of elastic solids. \textit{Proc R Soc Lond} : A324:301--313. doi: 10.1098/rspa.1971.0141

Maugis, D., \& Barquins, M. Fracture mechanics and adherence of viscoelastic
solids. In: Adhesion and adsorption of polymers. Springer, Boston, MA, 1980.
p. 203-277.

Muller, V. M. (1999). On the theory of pull-off of a viscoelastic sphere from
a flat surface. \textit{Journal of Adhesion Science and Technology}, 13(9), 999-1016

Persson, B. N. J., \& Brener, E. A. (2005). Crack propagation in viscoelastic
solids. \textit{Physical Review E}, 71(3), 036123.

Roberts, A. D. (1979). Looking at rubber adhesion. \textit{Rubber Chemistry
and Technology}, 52(1), 23-42.

Violano, G., \& Afferrante, L. (2019). Adhesion of compliant spheres: an
experimental investigation. Procedia Structural Integrity, 24, 251-258.

Williams, M. L.; Landel, R. F.; Ferry, J. D. The Temperature Dependence of
Relaxation Mechanisms in Amorphous Polymers and Other Glass-Forming Liquids.
\textit{Journal of the American Chemical Society }1955, 77, 3701-3707.

\end{document}